# Multi-seeded melt growth (MSMG) of bulk Y-Ba-Cu-O using thin-film seeds

T. Y. Li[1], C. L. Wang[1], L. J. Sun[1], S. B. Yan[1], L. Cheng[1], and X. Yao[1, 2, a)],

[1]Department of Physics, Key Laboratory of Artificial Structures & Quantum Control, Ministry of Education, [2]State Key Laboratory for Metal Matrix Composites,

Shanghai Jiao Tong University, 800 Dongchuan Road, Shanghai 200240, China

J. Xiong[3], and B. W. Tao[3],

[3]State Key Lab of Electronic Thin Films and Integrated Devices,

University of Electronic Science and Technology of China, 610054 Chengdu, P. R. China

J. Q.Feng[4]，X. Y. Xu[4]，C. S. Li[4]

[4]Northwest Institute for Nonferrous Metal Research Xi' an, Shaanxi, 710016 P. R. China

D. A. Cardwell[5]

[5]Department of Engineering, University of Cambridge, Madingley Road, Cambridge CB3 0HE, UK

a) Corresponding author: Prof. Xin Yao Tel. +86-21-54745772, Fax. +86-21-54745772

E-mail address: xyao@sjtu.edu.cn

**Abstract**

Y-Ba-Cu-O (YBCO) and Sm-Ba-Cu-O (SmBCO) thin films have been used for the first time as heterogeneous seeds to multi-seed successfully the melt growth of bulk YBCO in a multi-seeded melt growth (MSMG) process. The use of thin film seeds, which may be prepared with highly controlled orientation (i.e. with a well-defined a-b plane and precisely known a-direction), is based on their superheating properties and reduces significantly contamination of the bulk sample by the seed material. A variety of grain boundaries were obtained by varying the angle between the seeds. Microstructural studies indicate that the extent of residual melt deposited at the grain boundary decreases with increasing grain boundary contact angle. It is established that the growth front proceeds continuously at the (110)/(110) grain boundary without trapping liquid, which leads to the formation of a clean grain boundary.

**1. Introduction**

The top-seeded melt-growth (TSMG) method is used widely in the fabrication of single grain Y-Ba-Cu-O (YBCO) bulk superconductors. However, one of the major limitations of the TSMG technique is the long processing time due to the relatively low growth rate of the superconducting $YBa_2Cu_3O_7$ (Y-123) phase, particularly for the fabrication of large grains. This time-consuming procedure causes a variety of problems, such as self-induced crystallization and Ostwald ripening of non-superconducting $Y_2BaCuO_5$ (Y-211) particles at the growth temperature, which interfere with the growth process and degrade the flux pinning properties of the YBCO bulk grain.[1-2] As a result, increasing the growth rate of large, single grains is highly desirable for the growth of bulk samples with good superconducting properties.

Several techniques have been developed for the growth of large YBCO bulk superconductors over the past 20 years. Initially, the growth rate of the Y-123 phase was enhanced by a partial substitution of Y by rare earth (RE) elements, such as Nd, Sm and Pr, due to the increased RE solubility in the Ba-Cu-O liquid produced during the peritectic decomposition of YBCO.[3] Secondly, increased growth rate of single YBCO grains was achieved via the use of a large seed, which maximizes the size of the single domain.[4-6] Finally, the multi-seeding melt-growth (MSMG)[6-9] technique has been used to grow individual YBCO grains simultaneously from several seeds in order to reduce the growth time. Unfortunately, the MSMG method may result in the trapping of unreacted liquid phase material at grain boundaries, which can degrade the field trapping ability of the bulk sample.[7-11] The presence of unreacted liquid has been observed frequently at the (100)/(100) grain boundary, but rarely at the (110)/(110) grain boundary.[8-10]

SmBCO (Sm-123) and NdBCO (Nd-123) bulk and single crystal superconductors are used commonly as seeds for the growth of single grain YBCO due to their good lattice match and higher melting points than the Y-123 phase. However, a number of problems are associated with these seed materials in the MSMG method. Firstly, both SmBCO and NdBCO seeds introduce Nd and Sm into the bulk YBCO composition, which substitute onto the Ba site in the Y-123 phase and degrade significantly its superconducting properties (mainly in the vicinity of the seed). Secondly, the contact angle of grains cannot be controlled accurately in a multi-seeding process. In previous reports, YBCO/MgO thin films with high thermal stability have been demonstrated to act successfully as seeds for the growth of bulk YBCO.[12] The many desirable properties of YBCO thin films such as superheating, the induction of homo-epitaxial growth and their ability to be cleaved with a precise orientation makes them potentially suitable to overcome the major limitations of the multi-seeding process.

YBCO thin films were used in this study as homo-seeds in the MSMG process. Various grain boundaries were obtained by adjusting the orientation of the thin film seeds. The correlation between seed orientation, grain boundary contact angle and the presence residual liquid at grain boundaries has been studied.

## 2. Experimental Method

$YBa_2Cu_3O_{7-\delta}$ (Y-123) and $Y_2BaCuO_5$ (Y-211) powders were prepared by multiple calcination at 900 °C for 48 h and intermediate grinding process of a stoichiometric mixture of commercially available $Y_2O_3$ (99.99%), $BaCO_3$ (99.0%) and CuO (99.0%) precursor powders.

The Y-123 and Y-211 powders were mixed subsequently with a molar ratio of Y-123:Y-211=1:0.3. 0.3wt% of $CeO_2$ powder was then added to the precursor and mixed thoroughly using an agate mortar and pestle. The resultant powder was pressed uniaxially into pellets of 20 mm diameter and 8 mm thickness. Two YBCO film-seeds, were placed on the top surface of each precursor pellet with various angles Φ between their primary axes of between 0° and 90°, as shown schematically in Fig. 1. YBCO thin films were deposited by magnetron sputtering onto MgO (100) substrates of thickness 200 nm. The samples were deposited at 740 °C in an $Ar/O_2$ mixture by means of the inverted cylindrical target sputtering with a biaxially rotating substrate. (100)/(100) and (110)/(110) bulk grain boundaries were formed in this way for $\Phi = 0°$ and 90°, respectively, with intermediate grain boundary angles for values of $\Phi$ within this range. The process was performed in an air atmosphere. Each sample/seed arrangement was heated from room temperature to 900 °C in 4 h, and held at this temperature for 1.5 h. The samples were then heated to 1022 °C in 1.5 h, held at this temperature for 1.5 h, cooled rapidly to 1000°C in 15 minutes and then cooled slowly for 10 h at a rate of 0.4 °C/h. After that, the samples were furnace-cooled to room temperature. Finally, the small specimen that was cut from the bulk sample carefully was annealed in flowing oxygen gas at 475 °C for 150 h.

The fully processed samples were cut into smaller specimens and polished to investigate the grain boundary region using polarized light microscopy. The superconducting transition temperature $T_c$ was measured using a SQUID magnetometer (Quantum Design, MPMS-7).

## 3. Results and Discussion

Top views of the YBCO bulk samples fabricated by MSMG using two YBCO thin films are shown in Fig. 2 for (a) $\Phi = 0°$ and (b) 90°. Characteristic growth facet patterns can be observed in the single-domain regions, indicating that the grains are oriented with their c-axes parallel to the thickness of the pellet (i.e. into the plane of the paper). Fig.2 (c) shows the x-ray diffraction pattern of YBCO bulk with multi-seeds. It is obvious that this sample is highly oriented with the c-axis perpendicular to the surface, because only (0 0 l) diffraction peak were observed. It is apparent from the orientation of the seeds in this figure that (100)/(100) and (110)/(110) grain boundaries have formed with $\Phi = 0°$ and 90° in the center of each sample. Fig. 3 presents an XRD pole figure, showing 0° aligned four-fold symmetry on the YBCO film-seeds. It may be concluded, therefore, that the YBCO thin film seeds are oriented epitaxially with the [100] orientation of the MgO substrate (i.e. YBCO [100] $_{film}$ // MgO [100] $_{substrate}$), which results directly in the observed [100] orientation of the multi-seeded bulk YBCO sample (i.e. YBCO [100] $_{bulk}$ // YBCO [100] $_{film}$ // MgO [100] $_{substrate}$.). This is the first time that YBCO film-seeds have been used successfully in the multi-seeded melt growth of bulk YBCO. The use of thin film seeds offers a number of advantages over other, predominantly bulk, seeds. Firstly, YBCO thin films provide ideal homogeneous seeds in the MSMG process for the nucleation and growth of YBCO due to their identical lattice match with the bulk Y-123 phase. Secondly, hetero-seeds, such as SmBCO, NdBCO single crystals, which are used widely in the TSMG method, introduce impurity elements (Sm, Nd in particular) into the Y-123 phase during the growth process, which inevitably reduces $T_c$ of the fully processed sample in the vicinity of the seed, in particular.[13] The use of YBCO thin films as homo-seeds overcomes this problem directly. Thirdly, the growth of Y-123 thin films on single crystal MgO substrates of well-defined crystallographic orientation enables precise control of the orientation of the crystallographic c-axis of the Y-123 phase. In addition, the MgO single crystal substrate cleaves easily along the a/b- direction of the film, which enables the production of YBCO thin films of controlled orientation in regular, square planar geometries. This results in better control of the orientation of the seeded grain compared to that obtained with other (non thin film) seed materials, which characteristically have irregular geometries. Finally, a large number of relatively large areas YBCO films with well-defined c-axis orientation can be obtained readily from a single thin film, which is not the case for single crystal and bulk, melt processed seeds.

As seen form Fig.4, the temperature dependence of the magnetic susceptibility of the YBCO bulk was measured, a field of 10 Oe parallel to the c-axis of the sample (H//c) was applied. $\Delta T_c$ (the difference between 10% and 90% transition to the zero-field-cooled value ) is less than 0.5K. Obviously, the superconducting properties of YBCO bulk seeded by YBCO thin films are comparable to those seeded by other seed materials.

Figures 5 (a) and (b) show optical micrographs of the (100)/(100) and (110)/(110) grain boundaries of the YBCO bulk samples prepared by the MSMG process (indicated by arrows). It can be seen that significantly less residual melt is present at the grain boundary in part (b) of the figure than in part (a).

The presence of residual melt at YBCO grain boundaries has been attributed previously to the final composition and wetting angle of the solidifying melt at the position of the boundary.[9] The results of the present study, however, indicate clearly that the presence of residual melt is also determined by the growth orientation of impinging grains. The growth fronts of two impinging YBCO grains are parallel to each other during the formation of the (100)/(100) grain boundary, as shown schematically shown in Fig. 6. In this case, the individual grains at the position of the grain boundary grow in opposite directions. The growth of both grains is terminated abruptly at the grain boundary, resulting in the deposition of Ba-deficient residual melt at this position (which prevents the formation of the Y-123 phase).[14-15] This results in the formation of a 'dirty' grain boundary at the interface of impinging (100)/(100) grains. A consequence of using seeds that are not orientated in-plane is to nucleate grains that grow with non-parallel growth fronts, as shown schematically in Fig. 7 for the formation of a (110)/(110) grain boundary. In this case solidification continues at the position of the grain boundary following initial impingement of the individual grains without trapped melt, leading to the formation of a 'clean' grain boundary. This grain boundary is much less likely to form a weak link to the flow of current (and hence trapped field), therefore, and has clear potential to improve the large scale superconducting properties of large, multi-grain bulk YBCO samples.

The parameter of "d" is used represent the width of the grain boundary in order to analyze quantitatively the amount of residual melt (indicated by the dashed lines in Fig. 5). The (100)/(100) and (110)/(110) grain boundaries of the YBCO bulk samples in Fig. 5 ("d1" and "d2") are 16.4 μm and 5.7 μm, respectively, with the smaller value of d indicating the cleaner grain boundary.

It appears from this investigation that a clean boundary can be prepared by the use of thin film seeds of specific orientation. Fig. 8 shows schematically an arrangement of two film-seeds and impinging grains in a multi-seeded growth process. A bulk sample processed using a similar arrangement of thin film seeds was fabricated by the MSMG technique. The microstructures of this sample at positions Ⓐ and Ⓑ were examined by optical microscopy. No residual melt was observed at position Ⓐ, where the angle between two grains is relatively large, $\alpha > 90°$. In contrast, the residual melt exists along the grain boundary at position Ⓑ, where the angle between two grains is relatively small, $\beta < 90°$. It is clear, therefore, that the angle of impingement determines fundamentally the formation of a clean boundary in MSMG-processed YBCO bulk materials. A series of bulk samples with grain boundaries oriented at $\Phi = 0°$, $37°$, $60°$ and $90°$ were fabricated by the MSMG technique to investigate further the relationship between the contact angle of the grains and the formation of residual melt at the grain boundary. Each sample was polished and its microstructure in the vicinity of the grain boundary examined by polarized light microscopy. The widths, d, of the various grain boundaries are summarized in Table 1. These data show that the width of the grain boundary decreases as Φ is increased.

Figure 9 shows the variation of "1/d" (squares) with contact angle between two impinging grains. The variation of trapped magnetic field (circles) with contact angle, taken from reference,[16] is shown in the same figure. It shows that the trapped magnetics filed at grain boundary increases with the inceasing angle between impinging grains (Φ). There is a similar variation between "1/d" and "Φ", which indicates that (i) the amount of residual melt at grain boundarys decreases with the increment in the value of "Φ" and (ii) that cleaner grain boundaries result directly in higher trapped fields.

It is well-known that there are two characteristic types of crystallographic alignment in c-axis oriented YBCO thin films: 0° in-plane oriented (YBCO [100] $_{film}$ // MgO [100] $_{substrate}$) and 45° in-plane oriented (YBCO [110] $_{film}$ // MgO [100] $_{substrate}$). A film of the latter orientation (i.e. 45° in-plane), confirmed by the measured pole figure shown in Fig. 10, was used to grow bulk YBCO by the MSMG technique in a separate experiment. A photograph of the top surface of the resulting sample is shown in Fig. 11. The as-grown bulk superconductor has an orientation of YBCO [110] $_{bulk}$ // YBCO [110] $_{film}$ // MgO [100] $_{substrate}$. It is relatively easy to fabricate impinging grains with a clean (110)/(110) grain boundary using these particular YBCO thin-film seeds. Additionally, in this case, growth proceeds parallel to the [110] crystallographic direction of the grain due to the orientation of YBCO [110] $_{film}$ // MgO [100], which leads directly to a higher growth rate than that observed along the [100] direction. [17] This is an additional potential advantage of the MSMG process. Finally, in order to process multi-grain bulk YBCO under a higher $T_{max}$ (of 1070 °C) to broaden the growth window, SmBCO thin-films were used successfully to seed the growth of an YBCO bulk sample with a diameter of 42 mm, as shown in Fig. 12. $T_{max}$ is the highest temperature in the growth process, which results in complete decomposition of RE-123 phase. The processing time for this sample was reduced significantly (to 60 hours) compared with the convention TSMG process for a sample of this size.

## 4. Conclusions

Multi-grain bulk YBCO samples have been grown successfully using the MSMG melt processing technique for the first time using epitaxial YBCO thin-films. Various YBCO grain boundaries have been fabricated by adjusting the orientation of the thin film seeds. Clean grain boundaries were obtained in a bulk sample with a (110)/(110) grain boundary. The extent of residual melt at the grain boundary is observed to depend critically on the contact angle between two impinging grains, and may be optimized to increase the field trapping ability of the bulk sample. Finally, SmBCO thin films have also been used successfully, again for the first time, to broaden the growth window of YBCO and to grown a large diameter bulk sample.

**Acknowledgement**

The authors are grateful for financial support from the NSFC (grant No. 50872082), the MOST of China (863 project No. 2007AA03Z206)，Specialized Research Fund for the Doctoral Program of Higher Education and the Shanghai Science and Technology Committee of China（Grant No. 08dj1400203）. The authors would like to thank Professor Y Yoshida for providing a high quality SmBCO thin film.

Table list:

Table 1. Width of grain boundary for different grain contact angle.

| d (μm) | 16.4 | 10.4 | 7.2 | 5.7 |
|---|---|---|---|---|
| Φ (°) | 0 | 37 | 60 | 90 |

Figure list:

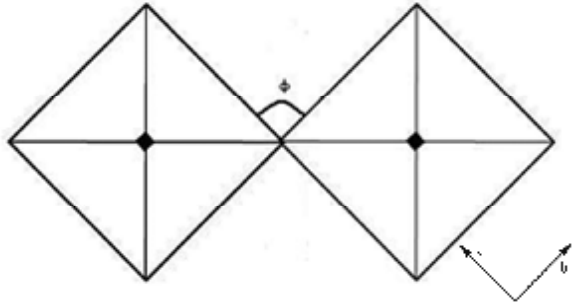

Fig. 1. The angle between the two grains was varied between 0° and 90°.

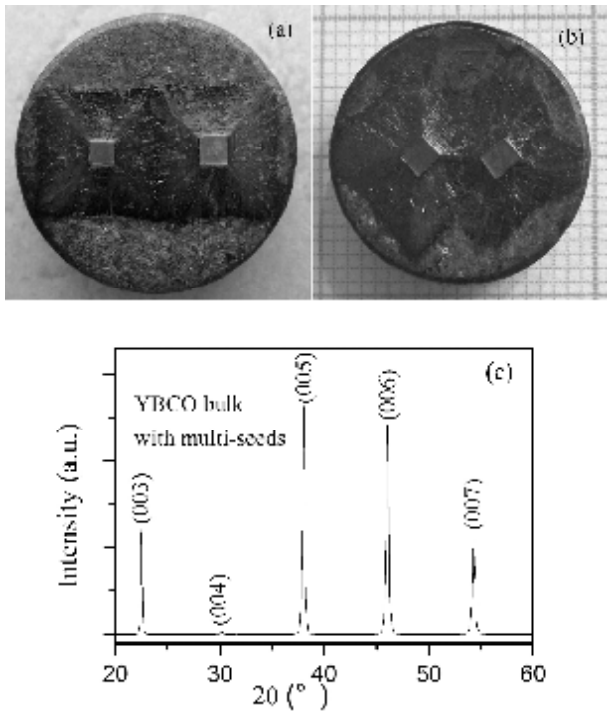

Fig. 2. Photographs of the top surfaces of YBCO bulk samples fabricated by the MSMG process with Φ = (a) 0° and (b) 90° using thin film seeds, (c) x-ray diffraction pattern of the YBCO bulk with multi-seeds.

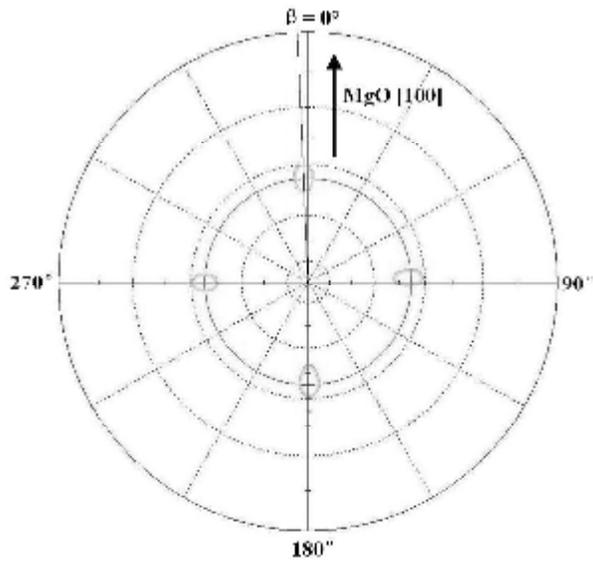

Fig. 3. XRD pole figure of an YBCO thin film with 0°in –plane orientation.

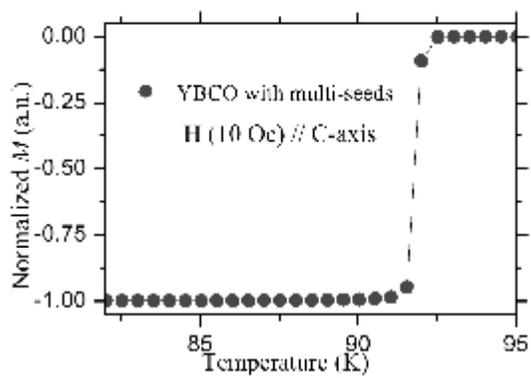

Fig.4 The temperature dependence of magnetization at a position of the YBCO bulk with multi-seeds, where is 5 mm away from the seed.

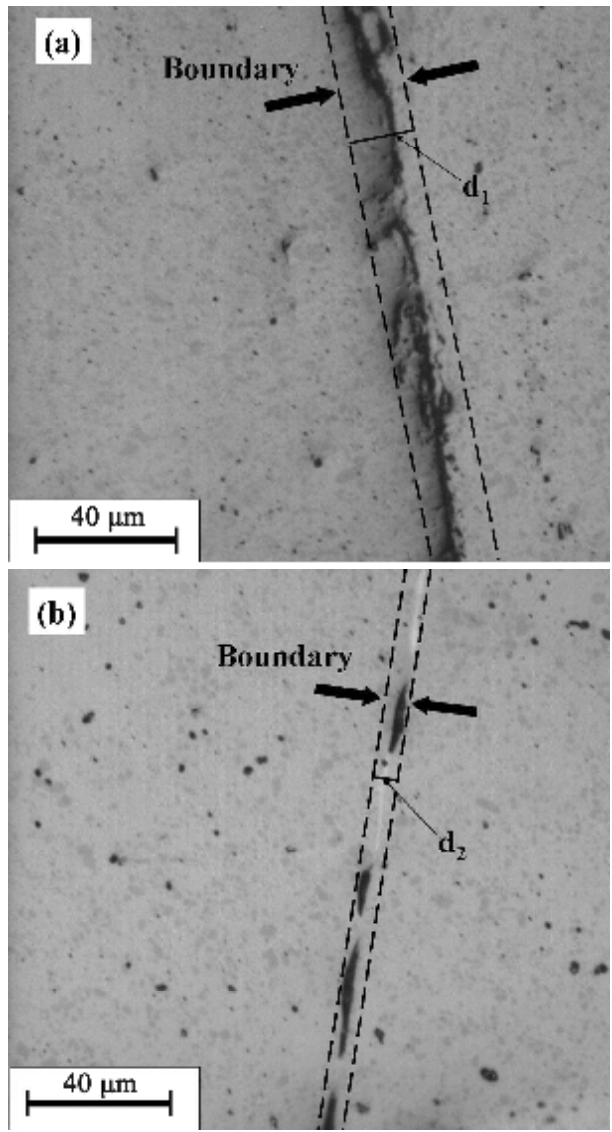

Fig. 5. Microstructure of the grain boundaries of the samples fabricated with Φ= (a) 0° and (b) 90°.

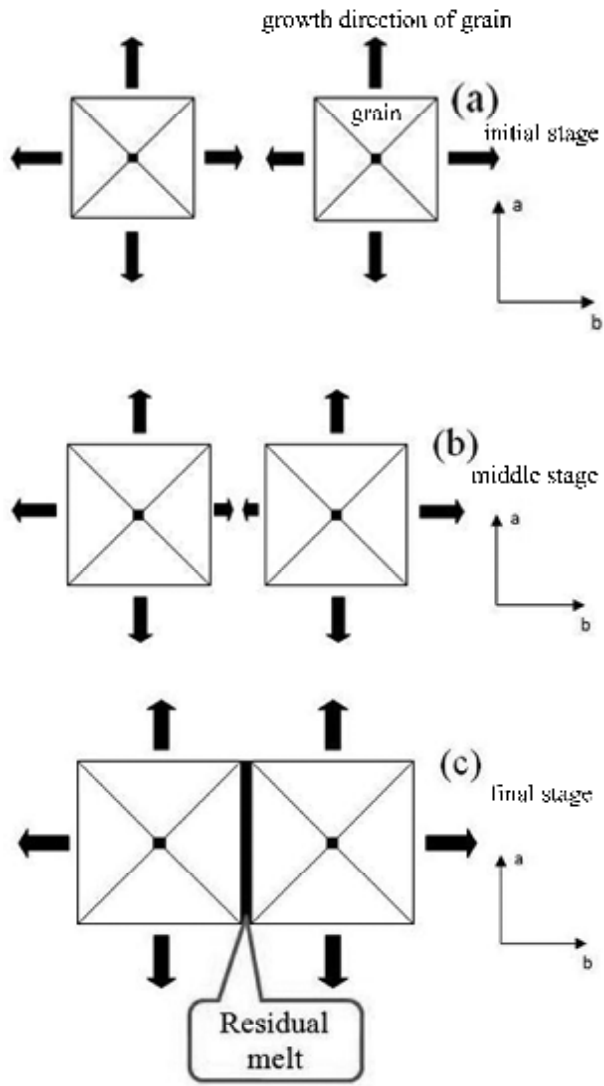

Fig. 6. Schematic illustration of the growth process and formation of the (100)/(100) grain boundary. The arrows correspond to the direction of motion of the growth front.

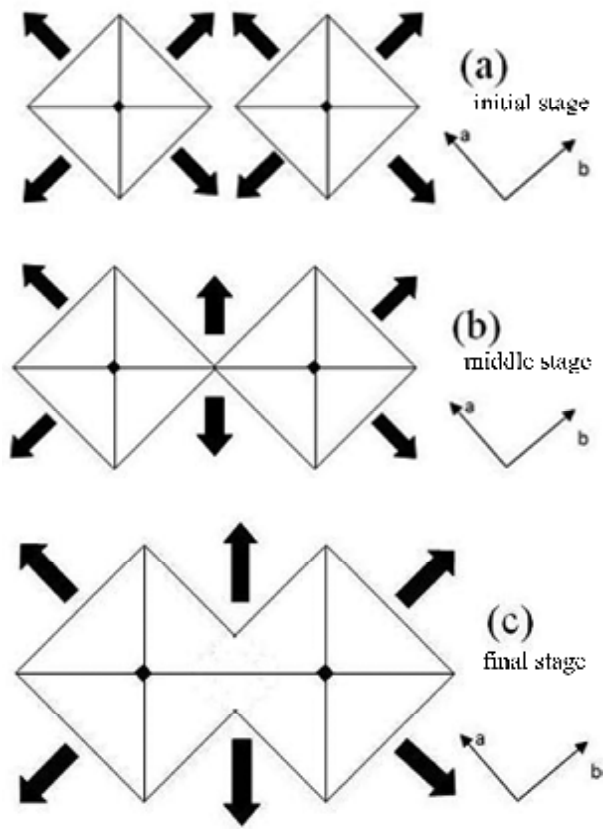

Fig. 7. Schematic illustration of the growth process and formation of the (110)/(110) grain boundary. The arrows correspond to the direction of motion of the growth front.

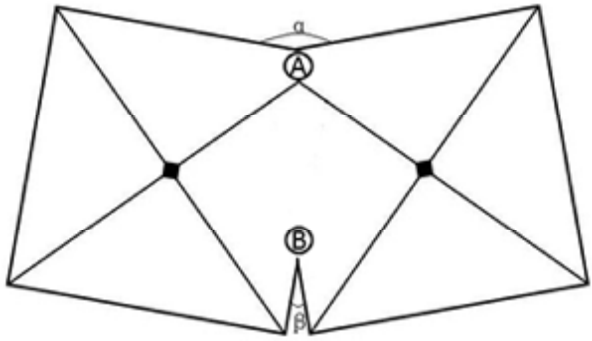

Fig. 8. Specific arrangement of grains with α > 90° and β < 90°.

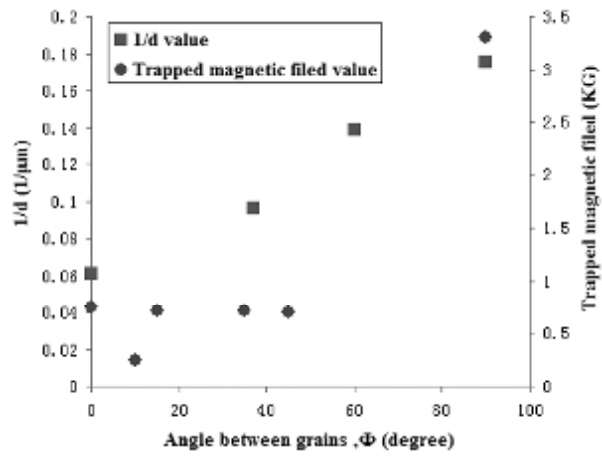

Fig. 9. The variation of "1/d" and trapped magnetic field as a funciton of Φ.

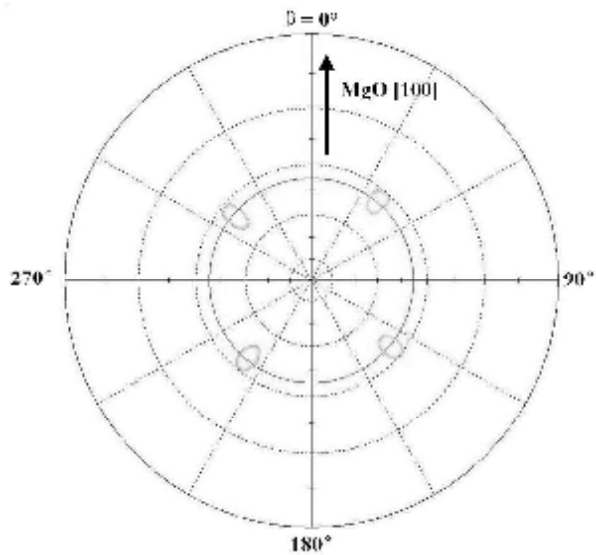

Fig.10. XRD pole figure of an YBCO thin film with 45°in –plane orientation.

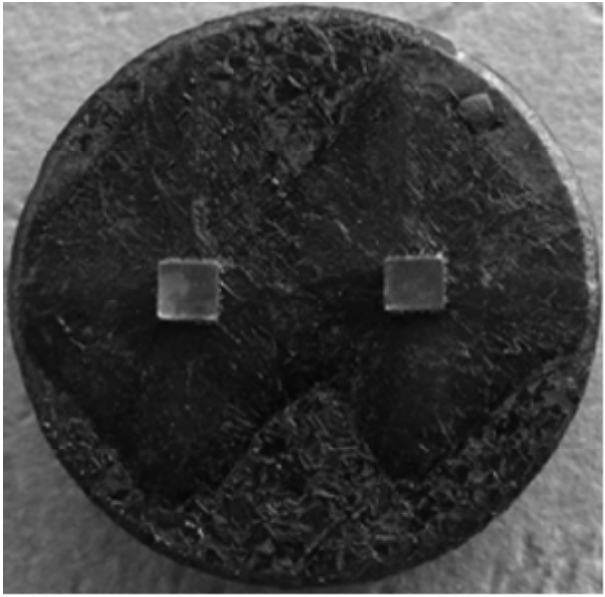

Fig. 11. Photograph of the top surface of a YBCO bulk superconductor processed by MSMG using 45° in–plane oriented YBCO thin film seeds.

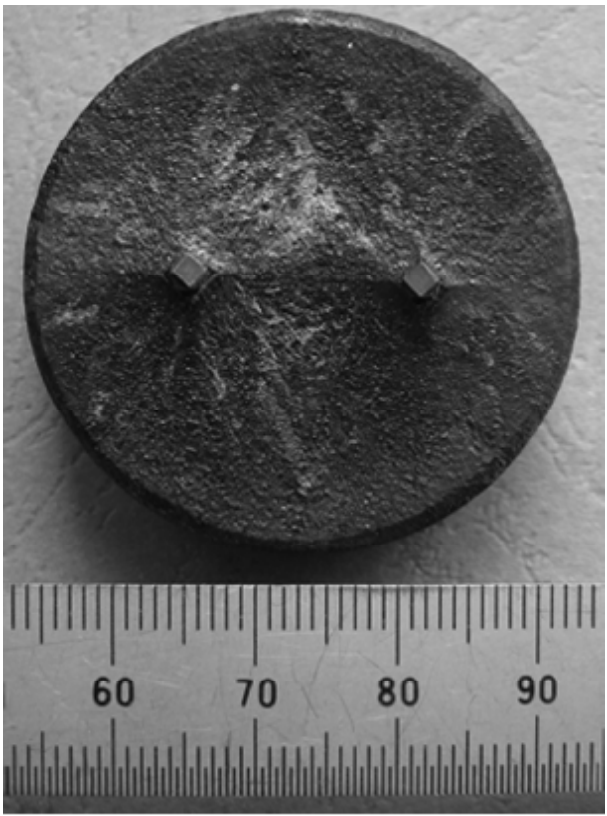

Fig. 12. Photograph of the top surface of a YBCO bulk superconductor processed using SmBCO thin film seeds.